\documentclass[runningheads]{llncs}
\pdfoutput=1
\usepackage{amssymb}
\usepackage{graphicx}
\usepackage{float}
\usepackage{url}
\urldef{\mailsa}\path|{alfred.hofmann, ursula.barth, ingrid.beyer,
christine.guenther,| \urldef{\mailsb}\path|frank.holzwarth,
anna.kramer, erika.siebert-cole, lncs}@springer.com|
\newcommand{\keywords}[1]{\par\addvspace\baselineskip
\noindent\keywordname\enspace\ignorespaces#1}

\begin{document}

\mainmatter  

\title{Long-Term Instrumental and Reconstructed Temperature Records
Contradict Anthropogenic Global Warming}
\author{Horst-Joachim L\"{u}decke}
\institute{EIKE, European Institute for Climate and Energy,\\
PO.Box 11011, 07722 Jena, GERMANY\\
\url{info@cfact-europe.org}\\\url{moluedecke@aol.com}\\---\\\begin{flushleft}Supplemented
version of the article published in Energy \& Environment, Vol. 22,
No. \nolinebreak 6 (Sept. 2011)
\end{flushleft}}

\toctitle{} \tocauthor{} \maketitle

\begin{abstract}
Monthly instrumental temperature records from 5 stations in the
northern hemisphere are analyzed, each of which is local and well
over 200 years in length, as well as two reconstructed long-range
yearly records - from a stalagmite and from tree rings that are
about 2000 years long. In the instrumental records, the steepest
100-year temperature fall happened in the 19$^{th}$ century and the
steepest rise in the 20$^{th}$ century, both events being of about
the same magnitude. Evaluation by the detrended fluctuation analysis
(DFA) yields Hurst exponents that are in good agreement with the
literature. DFA, Monte Carlo simulations, and synthetic records
reveal that both 100-year events have too small probabilities to be
natural fluctuations and, therefore, were caused by external trends.
In contrast to this, the reconstructed records show stronger
100-year rises and falls as quite common during the last 2000 years.
Consequently, their DFA evaluation reveals far greater Hurst
exponents. These results contradict the hypothesis of an unusual
(anthropogenic) global warming during the 20$^{th}$ century. The
cause of the different Hurst exponents for the instrumental and the
reconstructed temperature records is not known. As a hypothesis, the
sun's magnetic field, which is correlated with sunspot numbers, is
put forward as an explanation. The long-term low-frequency
fluctuations in sunspot numbers are not detectable by the DFA in the
monthly instrumental records, resulting in the common low Hurst
exponents. The same does not hold true for the 2000-year-long
reconstructed records, which explains both their higher Hurst
exponents and the higher probabilities of strong 100-year
temperature fluctuations. A long-term synthetic record that embodies
the reconstructed sunspot number fluctuations includes the different
Hurst exponents of both the instrumental and the reconstructed
records and, therefore, corroborates the conjecture.\keywords{
northern hemispheric 100-year temperature fluctuations, persistence
of temperature records, proxy temperatures from stalagmites and tree
rings, solar influence.}
\end{abstract}
\markboth{Horst-Joachim L\"{u}decke}{Long-term instrumental ....}
\section{Introduction}
It is widely assumed that the warming of the northern hemisphere
during the 20$^{th}$ century was anomalous. Greenhouse gases are
indeed an agent of warming, however, exactly by how much they
contributed to the 20th century rise in temperature and hence can be
linked to human influence, remains an issue that has yet to be
settled. The literature refers to this as the 'detection and
attribution problem' \cite{Barnett}, \cite{Hasselmann},
\cite{Hegerl}, \cite{Rybski2006}, \cite{Zwiers}, \cite{Zorita}. In
most of the papers discussing this problem, monthly local records of
up to 120 years were consulted. Some use a mixture of global and
local records. Up to 95 instrumental records going back 50 to 120
years from stations all over the globe were explored  but
indications of a warming of the atmosphere  could not be found in
the vast majority of cases \cite{Eichner2003}. A totality of 17
local records and 15 global records from all over the world were
examined for both the recent 50-year and the longer 100-year period,
as well as a further 13 local records from different parts of the
world for the last 50 years only \cite{Lennartz2009a}. The authors
found only weak support for the thesis that the warming trend in the
last 50 years changed its character compared with the first 50 years
of the 20$^{th}$ century.

Compared with the 20$^{th}$ century, a different picture arises at
the end of the 18$^{th}$ century because all reliable instrumental
temperatures for the northern hemisphere - there are no records for
the southern hemisphere that go far enough back in time - show a
100-year-long decline during the 19$^{th}$ century, turning into the
better-known, recent centennial rise at the end of it. Both rise and
fall are of similar magnitude. Therefore, if the 20$^{th}$ century
temperature rise is considered to be deterministic, the same should
apply to the fall in the 19$^{th}$ century. If we assume
anthropogenic CO$_2$ to be the agent behind the 20$^{th}$ century
rise, we face a problem when it comes to the 19$^{th}$ century.

Additional information about temperature trends of 100 years
duration is derived from reconstructed temperatures as 'proxies',
here from tree rings and stalagmites. Tree rings reflect the age of
a tree in yearly steps, and the thickness of the rings yields
information about the annual mean temperature \cite{Esper}. In the
thin layers of stalagmites, the age of the layer is determined by
the Th/U method, and the annual mean temperature of the layer is
deduced from the $^{18}$O value \cite{Mangini2005}. Two high-quality
proxies, one from tree rings and the other from a stalagmite, are
analyzed here. Both records go back about 2000 years and show during
this period even stronger 100-year-long temperature rises and falls
than prevailed in the 19$^{th}$ and 20$^{th}$ century. Hence, the
cause of both the 100-year-long temperature changes during the last
2000 years and the 20th century warming remains to be discovered.

Temperature series are persistent - other notations are 'long-term
correlated' or 'long-term memory' -, which is a well-known
phenomenon. A warm day is more likely to be followed by another warm
day than by a cold day, and vice versa. Short-term persistence of
weather states on a time scale of about one week is caused by
general weather situations. Longer-term persistence over several
weeks is generally caused by blocking situations that arise when a
high pressure system remains in place for many weeks. Persistence
over many months, seasons, years, decades, and even longer periods
is usually associated with anomaly patterns in sea surface
temperatures, and even with the influence of long-term variations in
the activity of the sun, but there is no universal explanation that
can be used in all causes \cite{Bunde2002}, \cite{Charney},
 \cite{Monetti}, \cite{Palmer}, \cite{Scafetta}.

In long-term correlated temperature series T$_i$, i = 1,....,N the
autocorrelation function
\begin{equation}
C(s) = \frac{1}{{\sigma _N^2 \left( {N - s} \right)}}\sum\limits_{i
= 1}^{N - s} {(T_i  - \left\langle T \right\rangle _N )(T_{i + s} }
- \left\langle T \right\rangle _N )
\end{equation}
decays with increasing s according to the power law
\begin{equation}
C(s) \sim s^{ - \gamma }
\end{equation}
where $\left\langle T \right\rangle _N = 1/N\sum\nolimits_{i = 1}^N
{T_i }$ is the average of the T$_i$, s is the time lag, and $\sigma
^2$ is the variance. Typical values for the exponent in Eq. (2) are
arranged between $0.4 < \gamma < 0.8$.

However, the most important characteristic of long-term correlated
records is that they include natural fluctuations that appear to be
trends caused by external impacts. Figure 1 depicts an example.
\begin{figure}[H]
\centering
\includegraphics[height=8cm]{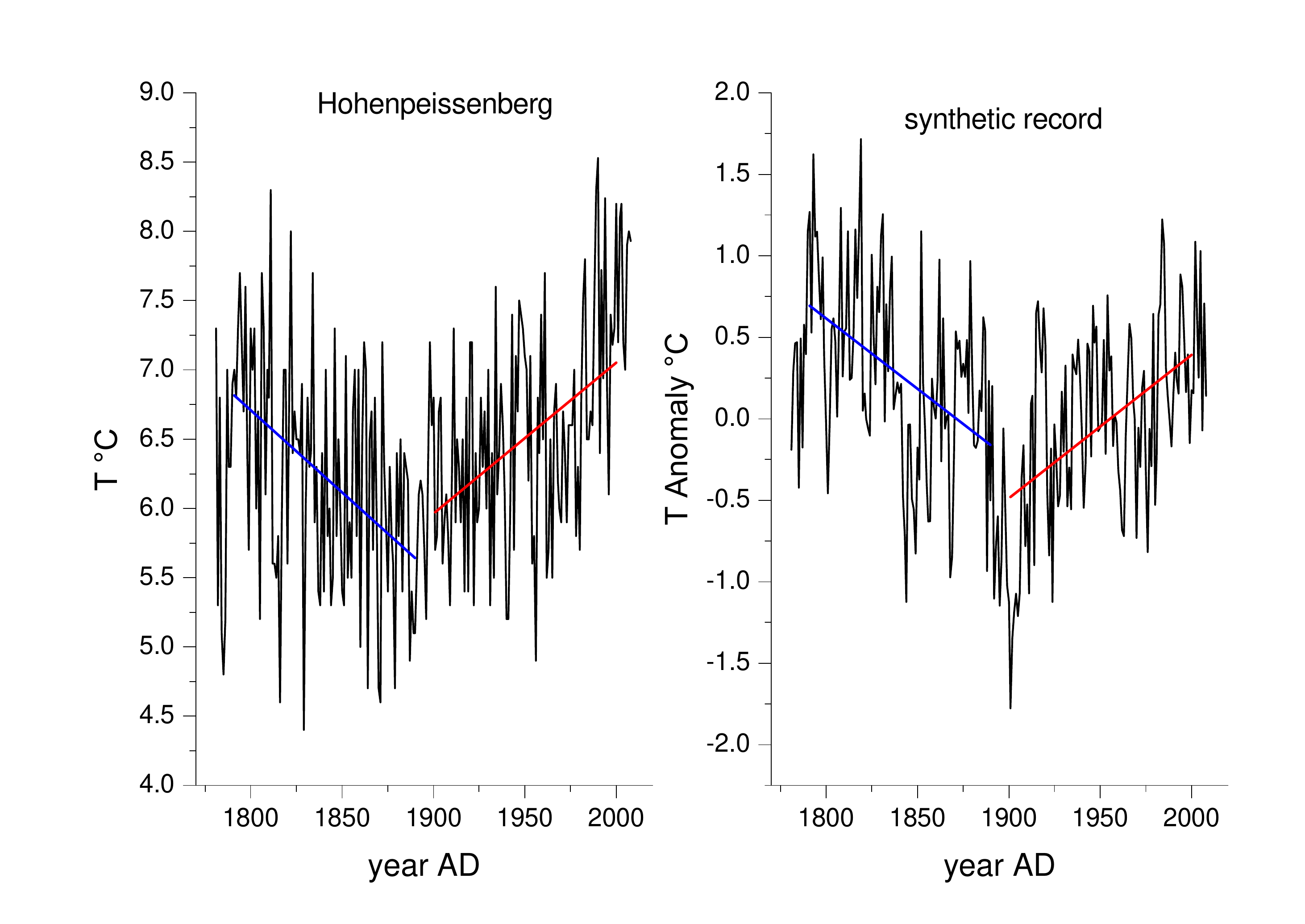}
\caption{(Color online) Comparison of the Hohenpeissenberg
temperature record for the period 1781-2008 (left panel) with a
synthetic record (right panel). The ostensible trends for the years
1789-1890 and 1901-2000 are similar in both temperature series.
However, the synthetic record is purely long-term correlated without
any deterministic trend.}
\end{figure}\newpage
The synthetic record shows apparently similar 100-year-trends as for
the Hohenpeissenberg series, but the synthetic data are only
long-term correlated. As a consequence, conventional methods based
on moving averages can no longer be used with persistent records to
separate deterministic trends from natural fluctuations.
Additionally, in persistent records often a clustering of extreme
events that is, in reality, quite natural will seem to be caused by
an external trend \cite{Bogachev2007}, \cite{Bunde2001},
\cite{Eichner2006}. Long-term correlations were not only detected in
temperature series, they are ubiquitous in nature at all time
scales. Among others, persistence was found in hydrological records,
economic records, physiological records, and earthquake events
\cite{Bogachev2009}, \cite{Eichner2006}, \cite{Ivanov},
\cite{Kantelhardt2001}, \cite{Kantelhardt2004},
\cite{Kantelhardt2006}, \cite{Kropp}, \cite{Livina2003},
\cite{Livina2005}, \cite{Lux}, \cite{Mantegna}, \cite{Rybski2008},
\cite{Rybski2009}. Therefore, the power law of Eq. (2) characterizes
the behaviour of persistent records quite generally.
\section{Methods: Detrended fluctuation analysis}
In this paper, linear regression lines are used for the analysis of
temperature series. A temperature rise or fall $\Delta _i$ is here
generally due to the (backward) linear regression line for the 100
annual temperatures T$_{i-99}$,....,T$_i$. Following
\cite{Lennartz2009a}, the dimensionless, relative temperature change
$\Delta _i/\sigma_t$ with $\sigma_t$ as the standard deviation
around the regression line is considered as the gauge for
temperature changes (the common standard deviation, which can be
affected by an external trend is not feasible here). The temperature
records that are analyzed in this article are available as
medium-range monthly instrumental records and as annual long-range
reconstructed records from tree rings and stalagmites. From the
monthly data, the seasonal effects have to be eliminated. At this,
for each calendar month the mean value of the whole record is
subtracted from the data and divided by the seasonal standard
deviation yielding the normalized record T$_i$, i=1,...,N that will
be applied in the further analysis.

In the following, the DFA (and FA) of temperature records is
described. If temperature records contain external trends, the
exponent $\gamma$ in Eq. (2) has not the correct value. Furthermore,
the direct calculation of C(s) by Eq. (1) is affected by side
effects, for instance, the shortness of the record, and by
deterministic trends \cite{Lennartz2009b}. Several methods to solve
these problems have been propagated, the fluctuation method
 (FA), the detrended fluctuation method (DFA) and
wavelet analysis \cite{Bunde2001}, \cite{Eichner2003},
\cite{Fraedrich}, \cite{Kantelhardt2001}, \cite{Kiraly},
\cite{Koscielny}, \cite{Pelletier}, \cite{Rybski2008},
\cite{Rybski2009}, \cite{Talkner}, \cite{Weber2001},
\cite{Yamasaki}. However, in particular the DFA yields correct
results if the record contains external trends. For a start, the
basic fluctuation analysis FA forms the following profiles as sums
over the normalized temperatures T$_i$:
\begin{equation}
Y_j  = \sum\limits_{i = 1}^j {T_i \;\;\;\;\;\;j =
1,....,N\;\;\;\;\;}
\end{equation}
Furthermore, it establishes the standard deviation F(s):
\begin{equation}
F(s)\; = \sqrt {\frac{1}{{(N - s)}}\sum\limits_{j = 1}^{N - s}
{\left( {Y_{j + s}  - Y_j } \right)^2 } }
\end{equation}
For long-term correlated data F(s) of Eq. (4) scales with a power
law
\begin{equation}
F(s) \sim s^\alpha
\end{equation}
The exponent $\alpha$ in Eq. (5) is linked to the power law of Eq.
(2) by
\begin{equation}
\begin{array}{l}
 \alpha  = 1 - \gamma /2\;\;\;\;\;\;\;for\;\;\;\;\;0 < \gamma  < 1 \\
 \alpha  = 0.5\;\;\;\;\;\;\;\;\;\;\;\;\;\;for\;\;\;\;\;\gamma  \ge 1 \\
 \end{array}
\end{equation} \cite{Kantelhardt2001}. The exponent $\alpha$ is
usually referred as Hurst exponent, because Hurst was the first who
detected long-term correlation in time series, but he used a
different method - the rescaled range (RS) analysis \cite{Hurst}.
Due to the already denoted limits of $\gamma$ together with Eq. (6),
$\alpha$ has the limits $0.5 \le \alpha  < 1$, where $\alpha$ = 0.5
stands for complete randomness, and $\alpha \approx 1$ for very
strong persistence. For $\alpha$ $>$ 1, the record becomes unsteady.

The detrended fluctuation analysis DFA can be seen as an advanced
FA, which automatically eliminates trends of the polynomial order
$\nu$ from the profile Y$_j$ and - because Y$_j$ in Eq. (3)
integrates the record $ \left\{ {T_i } \right\}$ - of polynomial
order ($\nu$ - 1) from the record itself. In DFA$_\nu$, the profile
Y$_j$ of Eq. (3) is divided into N$_s$ = int(N/s) non-overlapping
intervals of equal length s, and in each interval a polynomial
P$_{\nu}$ of order $\nu$ is evaluated, which provides the best fit
for the profile Y$_j$. Generally, a short segment at the end of the
profile remains. In order not to neglect this segment, one repeats
the same procedure from the other end of the record resulting in
2N$_s$ segments. In the next step the new profile Z$_j$ replaces
Y$_j$ of Eq. (3):
\begin{equation}
Z_j  = Y_j  - P_\nu
\end{equation}
Finally, with $G_k^2 (s) = \frac{1}{s}\sum\limits_{j = (k - 1)s +
1}^{ks} {(Z_j )^2 }$ the new F$_{\nu}$(s) replaces F(s) of Eq. (4):
\begin{equation}
F_\nu  (s)\; = \sqrt {\frac{1}{{2N_s }}\sum\limits_{k = 1}^{2N_s }
{G_k^2 (s)} } \;
\end{equation}
In this study, only DFA$_2$ is used. For purely long-term correlated
data, the fluctuation function F$_{\nu}$(s) scales like F(s) in Eq.
(5) with the same exponent $\alpha$. It is useful to depict
F$_{\nu}$(s) as function of s in a log-log diagram, which enables
the exponent $\alpha$ to be easily evaluated. It should be stressed
that Eq. (5) as well as the former Eq. (2) expresses the fact that a
consistent scaling behavior is induced by the numerous mechanisms
that generate persistence in temperature records. Consequently,
neither Eq. (2) nor a power law of the FA or DFA says anything about
its origin.

In principle, the power law of Eq. (5) does not depend on the time
scale of s - whether monthly or yearly. A minimum of $\nu$ + 2 data
points are needed for the fit procedure with polynomials of order
$\nu$, and for this reason, the F$_{\nu}$(s) graph begins with
s$_{min}$ = $\nu$ + 2. A minimum of about N = 600 data points T$_i$
are required to get reliable results from DFA. Consequently, monthly
not annual data are indispensable for the FA and DFA of instrumental
temperature records with 250 years length at the most. Furthermore,
reliable DFA results are confined to a maximum of s $\approx$ N/4
because otherwise the number of intervals N$_s$ = int(N/s) becomes
too small. Lastly, in F$_2$(s) the interval length s is too small
for about the first five s values, and due to numerical effects, the
Hurst exponent becomes slightly too high in this domain. This
reveals every double-logarithmic F$_2$(s) plot as a weak dip for a
few of the first s values (see Figure 4 in paragraph 4 as an
example). To summarize, DFA is reliable within
\begin{equation}
s_{\min }  + 5 \le s \le N/4\;\;\;\;\;\;\;and\;\;\;\;\;\;N \ge 600
\end{equation}
\section{Methods: Exceedance probabilities}
In the following, the probability  that in a long-term correlated
record with a Hurst exponent $\alpha$ the temperature change
$\Delta$ of a linear regression line over 100 years appears
naturally is of main interest. 'Natural' denotes that there is no
external trend in the record. Following  \cite{Lennartz2009a},
\cite{Lennartz2011}, \cite{Luedecke2011} W($\Delta / \sigma_t$,
$\alpha$) indicates for a fixed Hurst exponent $\alpha$ the
exceedance probability that values of $ \ge \Delta /\sigma _t$ over
100 years occur naturally. W($\Delta / \sigma_t$, $\alpha$) is not
restricted to positive $\Delta/\sigma_t$. If one begins, by
definition, with cooling, the exceedance probability converges to
the theoretical limit of W = 1 for extremely strong negative $\Delta
/ \sigma_t$ values and decreases with rising $\Delta / \sigma_t$. In
natural records we have equal fractions with warming and cooling,
hence for all $\alpha$ values W(0, $\alpha$) = 0.5 is valid. To give
further examples, W(-2, 0.7) = 0.9954, W(-0.8, 0.7) = 0.849, W(-0.3,
0.7) = 0.65, W(0, 0.7) = 0.5, W(0.3, 0.7) = 0.35, W(0.8, 0.7) =
0.151, and W(2, 0.7) = 0.0046 hold - by using later Eq. (11). The
general relation
\begin{equation}
W( - \Delta /\sigma _t ,\alpha )\; = \;1 - W(\Delta /\sigma _t
,\alpha)
\end{equation}
confirms congruent fractions with positive and negative $\Delta
/\sigma_t$ values in natural records. If values of W over 0.5 belong
to cooling, then, for instance, an exceedance probability of 0.9
would be as significant as of 0.1 for warming.

The latest method to evaluate the exceedance probabilities W($\Delta
/ \sigma_t$, $\alpha$) for natural records of a fixed $\alpha$ in a
systematical way is a combination of DFA, synthetic records, and
Monte Carlo simulations \cite{Lennartz2009a}, \cite{Lennartz2011}.
This method generates large - generally 2$^{21}$ = 2,097,152 data
points in length - long-term correlated synthetic records with this
Hurst exponent $\alpha$ and evaluates from these long records
\textit{n} short sub sequences for the period under consideration,
in our case 100 years, each with the same local $\alpha$ (in
general, the $\alpha$ values of sub sequences are similar but in
general not identical with the $\alpha$ of the long record). For
statistical reasons, the number \textit{n} must be sufficiently
large. In the stack of \textit{n} synthetic records, the number
\textit{z} of records with warming equal to or higher than $\Delta /
\sigma_t
> 0$ is counted. As a result, the fraction \textit{z/n} yields the exceedance
probability W($\Delta / \sigma_t$, $\alpha$) for positive $\Delta /
\sigma_t$ values in natural records. By systematically varying
$\Delta / \sigma_t$ and $\alpha$ the results can be expressed
empirically by the following analytic approximation for $\Delta /
\sigma_t > 0$ \cite{Lennartz2009a}:
\begin{equation}
W\left( {\Delta /\sigma _t ,\alpha } \right) = C \cdot \exp \left[ {
- B(\alpha )\left( {\Delta /\sigma _t  - 0.2} \right)} \right]
\end{equation}
For a period of 100 years the following parameters are valid: C =
1.15; B($\alpha$) = D$\alpha^{-\delta}$ with D = 1.26, $\delta$ =
2.5 \cite{Lennartz2009a}. Eq. (11) holds for $W \le 0.1$. For $W
> 0.1$, a continuation of Eq. (11) is yielded using the error
function \cite{Lennartz2009a}. For simplicity, this continuation is
included if Eq. (11) is referred to. The approximation of Eq. (11)
is completed for negative values of $\Delta / \sigma_t$ by Eq. (10).
It is demonstrated, for instance, if we were to choose $\Delta
/\sigma$ = 1 and $\alpha$ = 0.8, Eq. (11) yields the exceedance
probability W(1, 0.8) = 0.16 and furthermore by applying Eq. (10) of
W(-1, 0.8) = 0.84. Hence, the 16\% percentage of long-term
correlated or detrended real records with $\alpha$ = 0.8 and 100
years in length shows warming at magnitudes $\ge (\Delta /\sigma =
1)$. Further, the 84\% percentage of records shows a relative
temperature change $\ge (\Delta /\sigma = -1)$ and, conversely, the
16\% percentage of records a cooling at magnitudes $\le (\Delta
/\sigma = -1)$. W can increase by several orders of magnitude when
$\alpha$ increases. As a consequence, an increase of $\Delta /
\sigma_t$ $>$ 0, which is very unlikely for small $\alpha$ values,
becomes quite normal for larger ones. To give an example, $\alpha$ =
0.6 yields W(1.5, 0.6) = 0.003, but for $\alpha$ = 0.8 this value
increases to W(1.5, 0.8) = 0.66.

The use of the approximation in Eq. (11) for the analysis of real
temperature records that are detrended by DFA is based on the
hypothesis that real records are long-term correlated with an
additional trend. Furthermore, for a reliable analysis, the trend is
assumed to be linear. In order to distinguish between a natural
fluctuation and an external trend, a confidence level ''a" is
needed. Consequently, the domain $0 < W < a$ of warming and $(1 - a)
< W < 1$ of cooling is considered to include external trends. On the
other hand, relative trends in the (1 - 2a) confidence interval $a
\le W \le (1 - a)$ are regarded as natural. It is defined - to some
extent subjectively - the confidence level of 0.025, which
corresponds to a confidence interval of (1-2a) = 0.95 in accordance
with general practice.
\section{Application: Medium-range monthly instrumental temperature records}
Medium-range monthly instrumental records going back more than 200
years exist only for stations in the northern hemisphere. But even
here they are not abundant. The places with reliable monthly
temperature series number five in all: Prague, Hohenpeissenberg,
Vienna, Paris, and Munich, and they have some of the longest, most
reliable (and available) instrumental temperature records to be had
anywhere. In Table 1 their characteristics are cited. They go back
for somewhat more than 220 years,  with the earliest being 1757 for
Paris and the latest starting in 1781 in Hohenpeissenberg and Munich
and can be accessed at \cite{GISS}, \cite{klementinum},
\cite{NOAA2010}, \cite{wetterzentrale}. The related temperature
courses are shown in Figure 2.
\newline\newline
\small{
\parbox[]{15cm}{
\hspace{0cm}\begin{tabular}{p{3.3cm} p{2.5cm} p{6cm}} \hline\hline
Station &  Period & Data consistency\\
\hline
Hohenpeissenberg & 1781-2008 & complete \\
Munich-Riem & 1781-2009 & 1992-2009 data from Munich airport \\
Prague-Klementinum  &  1770-2009 & complete \\
Paris-le-Bourget & 1757-2008 & 1994-2009 data from Paris
waterworks \\
Vienna & 1775-2008 & complete \\
\hline
\end{tabular}}\newline\newline
Table 1: Medium-range monthly temperature records analyzed in this
study} \normalsize{}
\begin{figure}[H]
\centering
\includegraphics[height=7.5cm]{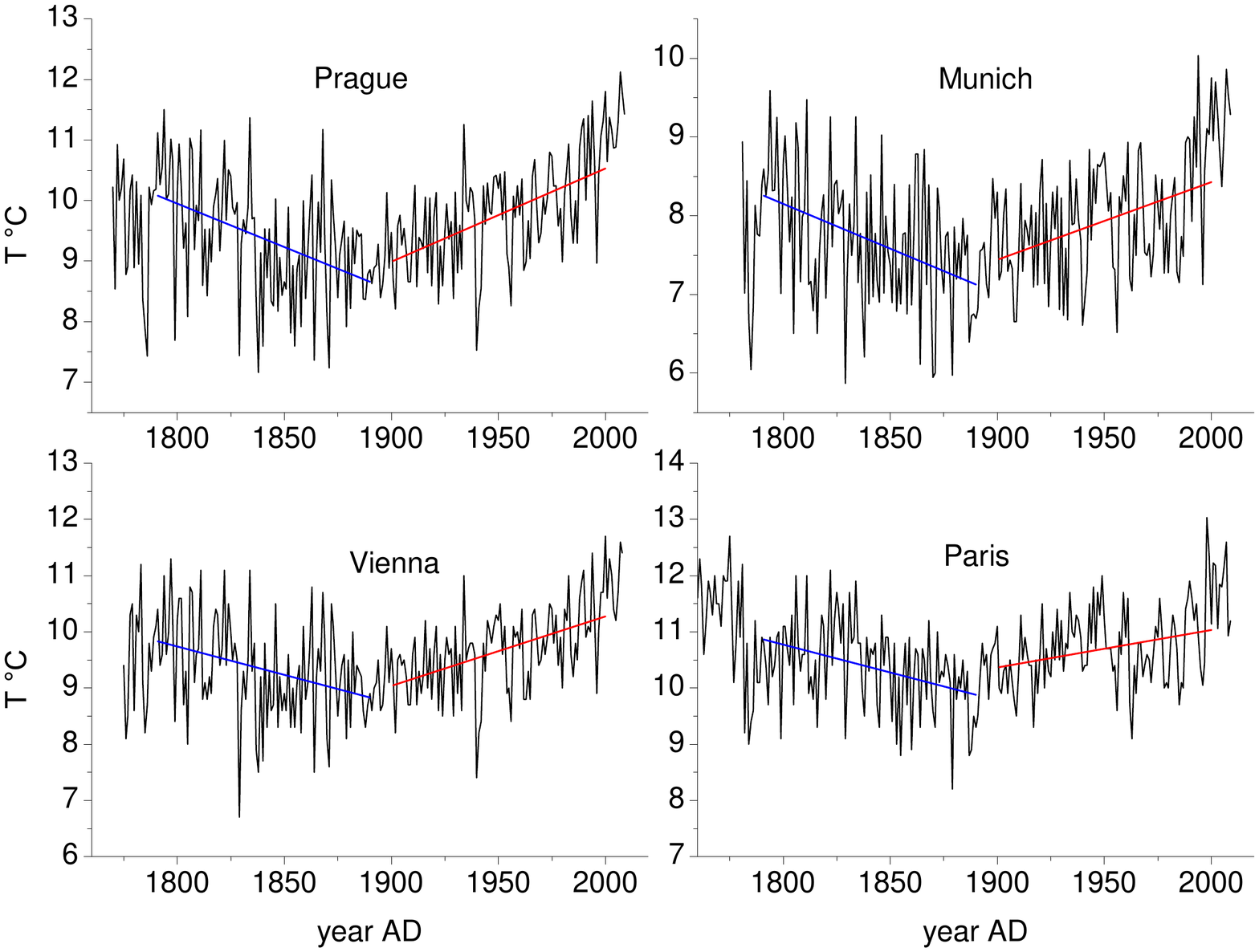}
\caption{(Color online) Temperature courses for Prague, Munich,
Vienna, and Paris with 100-year linear regression lines for
1791-1890 and 1901-2000 (the temperature course of Hohenpeissenberg
was already depicted in Figure 1 of paragraph 1).}
\end{figure}
The periods 1791-1890 and 1901-2000 were selected as having for all
stations roughly the steepest temperature fall and temperature rise
during the last 220 years. All the records show the same general
behavior: a temperature drop in the period 1791-1890 and a rise for
1901-2000 with both roughly the same magnitude.

The following Figure 3 depicts the values of $\Delta_i / \sigma_t$
with $\Delta_i$ as the temperature difference  due to the 100-year
linear regression line trough the data points T$_{i-99}$,....,T$_i$
and $\sigma_t$ as the appropriate standard deviation around the
line. The yearly indices i are moving earliest from the year (1757 +
99) until latest 2008 or 2009 (see Table 1). Die dashed line
$\Delta_i / \sigma_t$ = 2 in Figure 3 is for comparison with later
Figure \nolinebreak 6, which will reveal that the relative 100-year
temperature rises $\Delta_i / \sigma_t$ in the last 2000 years were
often far stronger than the appropriate rises of the 20$^{th}$
century.
\begin{figure}[H]
\centering
\includegraphics[height=7cm]{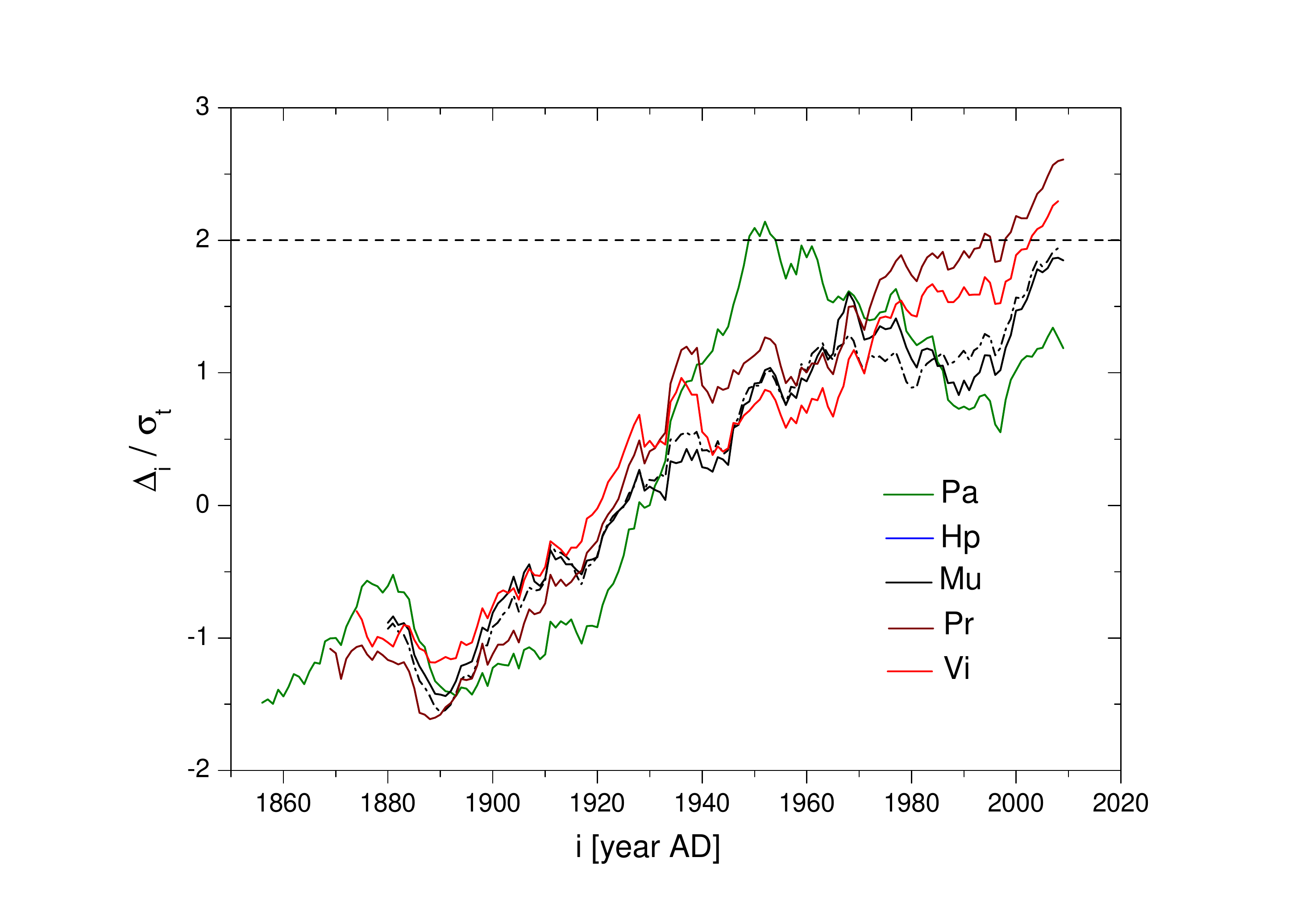}
\caption{(Color online) $\Delta_i / \sigma_t$ against years i, with
$\Delta_i$ as the rises and falls of the (backward) 100-year linear
regression lines through the temperatures T$_{i-99}$,....,T$_i$ and
$\sigma_t$ as the standard deviation around the lines, for the
medium-range series (top to bottom) Pa = Paris, Hp =
Hohenpeissenberg, Mu = Munich, Pr = Prague,  and Vi = Vienna. With
few exceptions, $\Delta_i / \sigma_t$ $\le$ 2 holds for all years
$\le$ 2000 (dashed line).}
\end{figure}
Figure 4 depicts the F(s) graphs of the FA in Eq. (4) and the
F$_2$(s) graphs of the DFA2 in Eq. (8) for the instrumental series.
All the graphs have a common shape. The numerical effect of the
somewhat greater Hurst exponents on a few of the first s values has
already been discussed in paragraph 2. A slight F$_2$(s) deviation
for the DFA2  graph of Paris shows a weak cross-over at about 600
months (50 years). Figure 4 reveals that $\alpha  > \alpha _2$ holds
for all records, which is an indication that linear trends in the
records have been eliminated by DFA2. Table 2 shows the appropriate
numerical results for $\alpha$ from the fluctuation analysis (FA)
and for $\alpha_2$ from the detrended fluctuation analysis (DFA2).
With the exception of Munich and Hohenpeissenberg in the period
1791-1890, Table 2 exhibits that for all records and all periods,
$\alpha  > \alpha_2$ is valid, indicating that linear trends have
been removed from the records by DFA2. The values of $\Delta_i$,
$\Delta_i / \sigma_t$, and $\sigma_t$ are cited in Table 3 for the
periods 1791-1890 and 1901-2000. Further, the exceedance
probabilities (100 - W) for cooling and W for warming (in percent)
derived from Eqs. (10)-(11) are given.

Within the already noted confidence interval of 95\%, the exceedance
probabilities in column 5 and column 9 of Table 3 are arranged in
the trend domain of W $ > $ 97.5\%, i.e. (100 - W) $<$ 2.5\% for
cooling and W $ < $ 2.5\% for warming. This holds for both
centennial occurrences and, therefore, both have to be judged as
nonnatural - with the only exception of Vienna for the period
1791-1890.\newline\newline \small{
\parbox[]{15cm}{
\hspace{0cm}\begin{tabular}{p{1.9cm} p{1.6cm} p{1.6cm} p{1.6cm}
p{1.6cm} p{1.6cm} p{1.6cm} }\hline\hline
 & $\alpha$ & $\alpha_2$ & $\alpha$ & $\alpha_2$ & $\alpha$ & $\alpha_2$ \\
&  1791- & 1791- & 1791- & 1791- & 1901- & 1901- \\
&  2000 & 2000 & 1890 & 1890 & 2000 & 2000 \\
\hline
Munich & 0.69 & 0.63 & 0.63 & 0.66 & 0.66 & 0.62 \\
Hohenpeis. & 0.67 & 0.60 & 0.61 & 0.63 & 0.67 & 0.59 \\
Prague  &  0.75 & 0.61 & 0.74 & 0.66 & 0.71 & 0.63 \\
Paris & 0.71 & 0.57 & 0.71 & 0.61 & 0.63 & 0.52 \\
Vienna & 0.72 & 0.59 & 0.67 & 0.66 & 0.69 & 0.59 \\
\hline
\end{tabular}}\newline\newline
Table 2: Hurst exponents $\alpha$  of the FA and $\alpha_2$ of the
DFA2 with the applied time periods. The appropriate graphs are
depicted in Figure 4.}\normalsize{}
\begin{figure}[H]
\centering
\includegraphics[height=7.2cm]{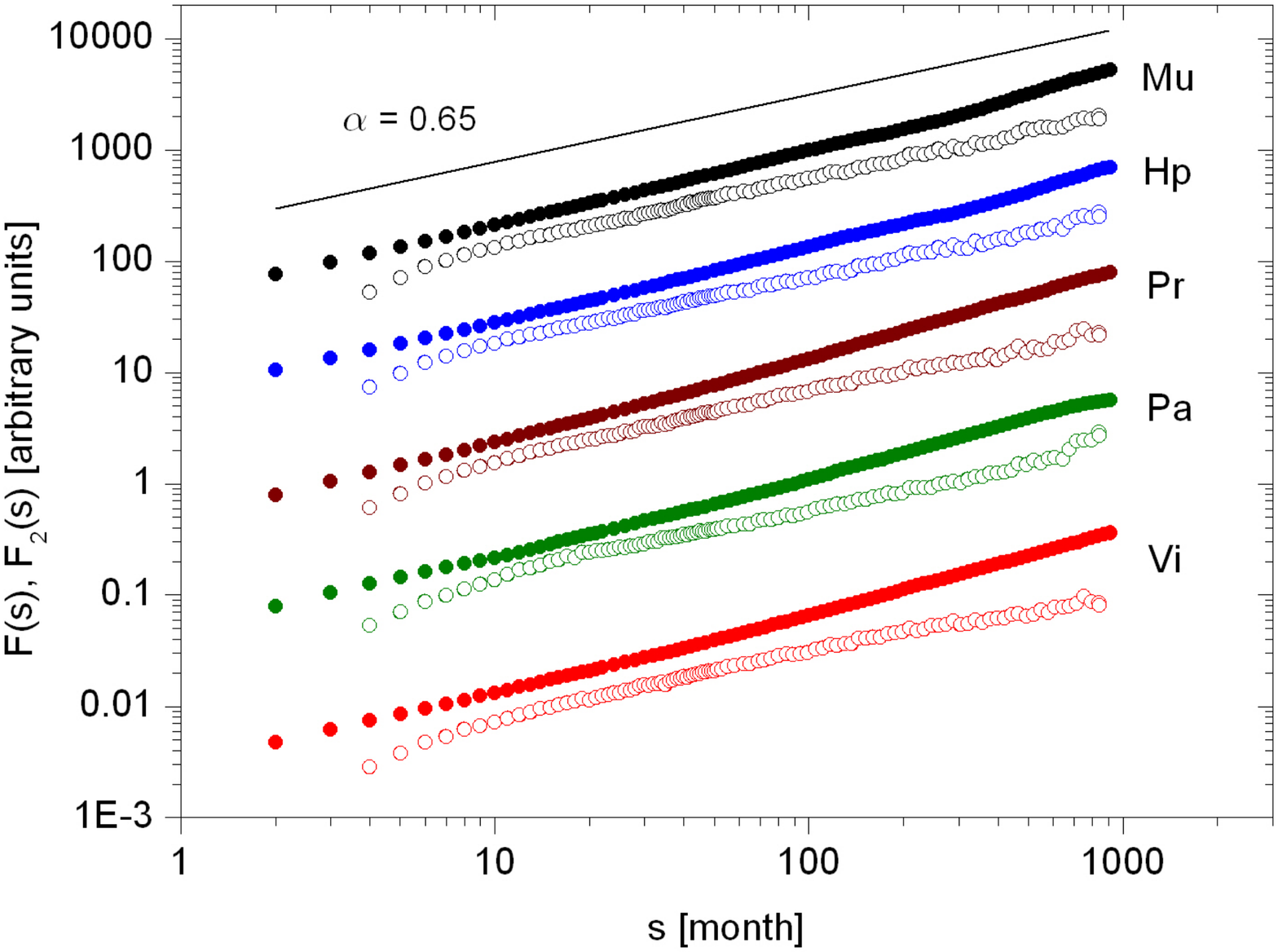}
\caption{(Color online) F(s) of the FA, shown as filled circles and
F$_2$(s) of the DFA2, shown as open circles for the records (top to
bottom) Mu = Munich (black), Hp = Hohenpeissenberg (blue), Pr =
Prague (brown), Pa = Paris (green), and Vi = Vienna (red) for the
period 1791-2000 in a double-logarithmic plot, and with the graphs
shifted for clarity. Obviously, all records show persistence over
more than 50 years. The related Hurst exponents $\alpha$ derived
from FA and $\alpha_2$ from DFA2 are cited in Table 2. The
appearance of $\alpha  > \alpha _2$
 in all records indicates that DFA2
has removed linear trends from all of them. The line with $\alpha$ =
0.65 is for comparison. }
\end{figure}
We do not know what caused these two centennial occurrences in
opposite direction. Consequently, if we would suppose anthropogenic
CO$_2$ as an agent for the recent warming we are faced with the
problem when we look for the agent of the cooling in the period
1791-1890. A further aspect worth mentioning is that the standard
deviations for 1791-1890 are generally higher than for 1901-2000.
The reason for this difference is also not known.

In particular for the last 50 years, instrumental temperatures in
populated areas are subject to the urban heat island effect (UHI)
and further warm bias. However, no such effects are considered here.
It should be emphasized that taking warm bias into account would
enforce the sought-after unbiased temperature falls for the 19th
century and have a dampening effect on the unbiased 20th century
rise caused by anthropogenic green house
gases.\newline\newline\small{
\parbox[]{15cm}{
\hspace{0cm}\begin{tabular}{p{1.5cm} p{1.1cm} p{1.1cm} p{1.1cm}
p{2cm} p{1.1cm} p{1.1cm} p{1.1cm} p{1.1cm}}\hline\hline
 & $\Delta_{1890}$ & $\Delta_{1890}/$ & $\sigma_t$
 & 100-W$_{1890}$ & $\Delta_{2000}$ & $\Delta_{2000}$/
 & $\sigma_t$ & W$_{2000}$ \\
& [$^0$C] & $\sigma_t$ & [$^0$C]
 & [\%] & [$^0$C] & $\sigma_t$
 & [$^0$C] & [\%]\\
\hline
Munich & -1.14 & -1.43 & 0.79 & 1.4 & 0.99 & 1.46 & 0.67 & 0.6 \\
Hohenp. & -1.19 & -1.54 & 0.76 & 0.5 & 1.09 & 1.56 & 0.69 & 0.2 \\
Prague  &  -1.44 & -1.57 & 0.91 & 0.9 & 1.54 & 2.17 & 0.71 & $< 0.1$ \\
Paris & -0.99 & -1.36 & 0.72 & 0.7 & 0.67 & 1.03 & 0.64 & 0.5 \\
Vienna & -1.01 & -1.17 & 0.86 & 3.6 & 1.23 & 1.89 & 0.65 & $<
0.1$ \\
\hline
\end{tabular}}
\newline\newline
Table 3: $\Delta_{1890}$ in column 2 denotes the temperature fall of
the linear regression line for the 100-year period 1791-1890. The
columns 3, 4, and 5 show $\Delta_{1890} / \sigma_t$, $\sigma_t$ as
the standard deviation around a regression line, and (100 -
W$_{1890}$) with W$_{1890}$ as the exceedance probability in percent
that temperatures during the period 1791-1890 with values $\ge
\Delta / \sigma_t$ occur in the DFA-detrended record naturally.
Columns 6 - 9 give the analogous results for the 100-year
temperature rise of the period 1901-2000.}\normalsize{}
\section{Application: Reconstructed long-range annual records}
Two long-range annual records, which are depicted in Figure 5 were
considered in this letter: a stack about tree rings and further
biological proxies \cite{NOAA2010}, referred to here as MOB, and the
SPA12 record of stalagmites \cite{Mangini2005}. MOB covers the
period 0 - 1979 AD and is a composition derived from wavelet
analyses of 11 low-resolution series - ice from boreholes,
Foraminifera and stalagmites - and 7 tree ring proxies with annual
or decadal time resolutions found in different sites of the northern
hemisphere \cite{Moberg}. MOB is not local and therefore an
exception to all other records analyzed in this paper. In the case
of SPA12, the dated isotopic composition of a stalagmite from the
Spannagel Cave at 2347 meter above sea level in the Central Alps
near Innsbruck was translated into a highly resolved record of
temperatures during the past 2000 years \cite{Mangini2005}. SPA12
goes from -90 to 1935 AD, and the average time resolution is
slightly over one year, with a minimum of 0.56-year- and a maximum
of 10-year steps (10-year steps in only 9 cases). For the purpose of
the DFA analysis, SPA12 was standardized here to one-year steps by
Spline interpolation. After this standardization no differences to
the original record could be detected by eye, and no differences in
the later DFA analysis were found by using interpolations with
different Spline functions. The variance \textit{var} = 0.048 of MOB
is smaller than the variance \textit{var} = 0.345 of SPA12. As a
further comparison, the annual mean temperature inside the Spannagel
cave was instrumentally evaluated as 1.8 $^0$C for the year 2003
alone \cite{Mangini2005}.
\begin{figure}[H]
\centering
\includegraphics[height=7.8cm]{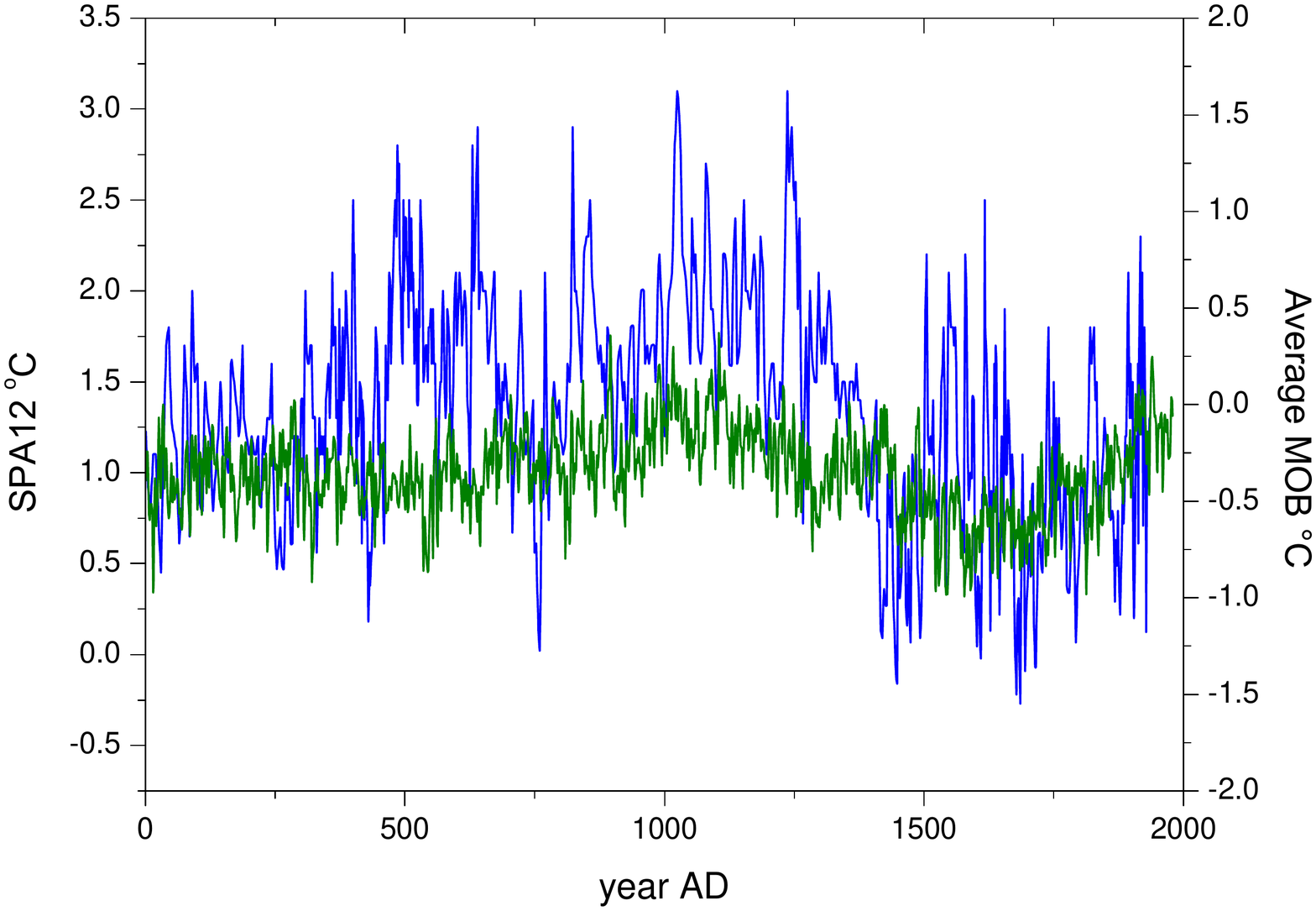}
\caption{(Color online) Reconstructed long-range annual records,
SPA12 (blue) from a stalagmite and MOB (green) from a stack of tree
rings in the northern hemisphere and further biological proxies (as
an anomaly). The variance of MOB is roughly seven times smaller than
for SPA12. In both records, the medieval warm period and the Maunder
Minimum can be clearly identified.}
\end{figure}
A comparison of tree rings and additional biological proxies with
SPA12 is discussed in \cite{Mangini2005}, \cite{Mangini2007}. The
uncertainty of the SPA12 temperatures is indicated to be $ \pm 0.3$
$^0$C in \cite{Mangini2005}, and $ \pm 0.4$ $^0$C for MOB in
\cite{Moberg}. The most conspicuous difference between stalagmite
and tree ring proxies has to do with the variance. It is reported
that smoothed MOB and SPA12 curves are also in accordance with other
biological proxies and with a combined temperature reconstruction of
Greenland ice cores and tree rings from Scandinavia and Siberia
\cite{Luterbacher}. In \cite{Mangini2005}, a comparison of the
smoothed SPA12 record with a sea surface temperature (SST) model
curve of the Bermuda Rise yielded by \cite{Loehle} is depicted that
also shows reasonable accordance. Further, a comparison of SPA12
with an annual winter curve of the Alps is given in
\cite{Luterbacher}. However, it is known that the thickness of tree
rings depends not only on the annual mean temperatures, but also on
precipitation. Similar problems are reported for stalagmites
\cite{Mangini2005}, \cite{Mangini2007}. As a consequence, short time
periods show a lack of congruence between MOB and SPA12 as well as
between the different tree ring proxies themselves
\cite{Rybski2006}.

Figure 6  shows  the 100-year (left panel) and the 500-year (right
panel) events $\Delta_i / \sigma_t$, with $\Delta_i$ as the rise or
fall of linear regression lines on the time intervals [i-99, i]
resp. [i-499, i] for SPA12 and MOB. In the 100-year cases, $\Delta_i
/ \sigma_t$ values of $ > 2$ and $< -2$ are commonplace during the
last 2000 years (see dashed line for comparison). Conversely, all
instrumental medium-range records before the year 2000 have $\left|
{\Delta _i /\sigma _t } \right|$ values of about 1.5, with a maximum
for Prague as the single exception (see Figure 3 and Table 3). This
in itself indicates that neither the 19th century fall nor the 20th
century rise in temperature are exceptional in the long run.
Concerning the 500-year graphs, the right panel of Figure
\nolinebreak 6 depicts additionally the 500-year
$\Delta$SSN$_i$/$\sigma_t$ of the annual SSN sunspot number record
for comparison (see paragraph 6 for relevance).
\begin{figure}[H]
\centering
\includegraphics[height=7.8cm]{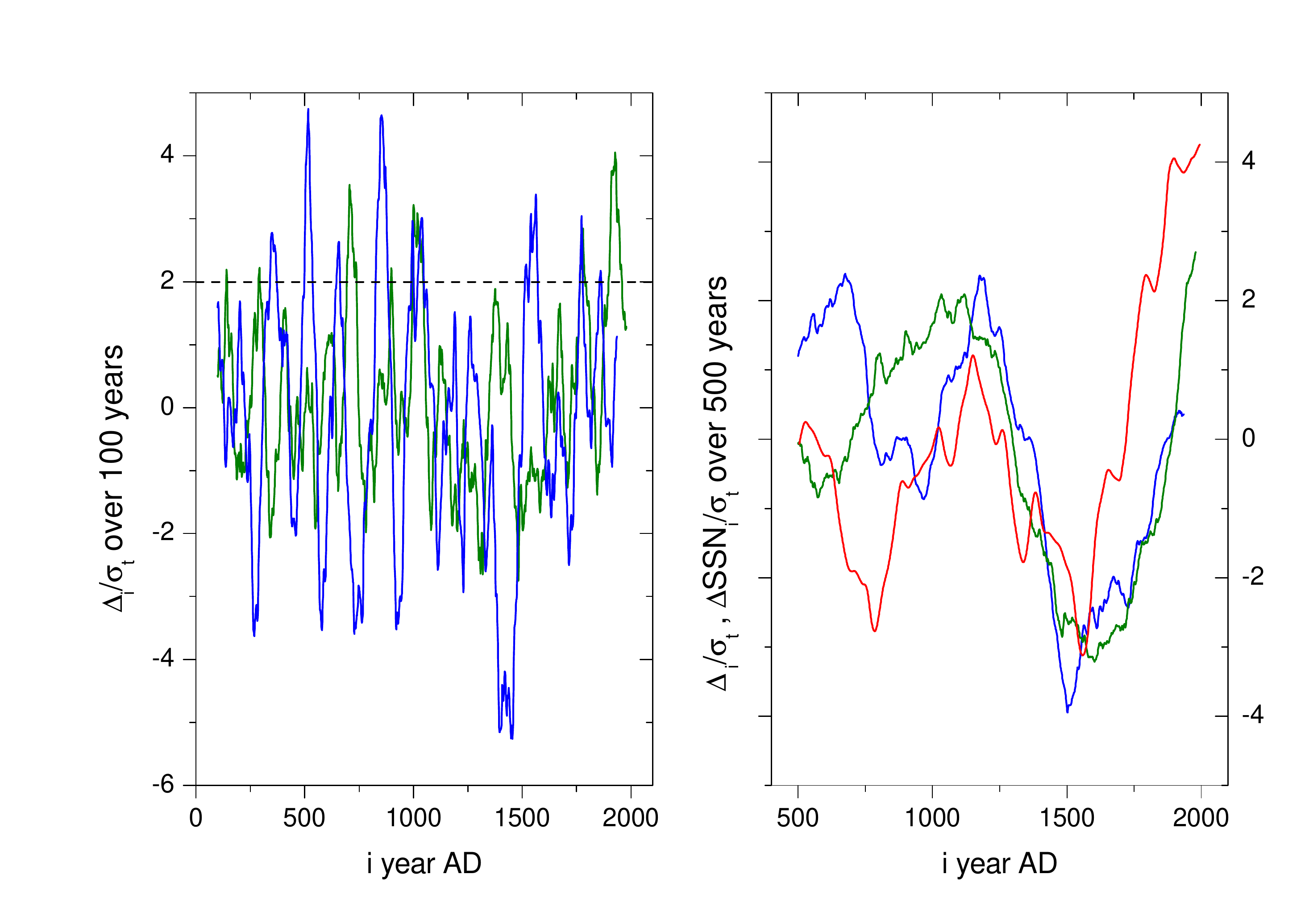}
\caption{(Color online) Left panel:  $\Delta_i / \sigma_t$, with
$\Delta_i$ as the rise or fall of 100-year linear regression lines
for the temperatures on the time intervals [i-99, i] and $\sigma_t$
as the standard deviation around the lines, for SPA12 (blue) and MOB
(green). Values of $\Delta_i / \sigma_t$ $ > $ 2 are quite common
during the last 2000 years (dashed line for comparison). Right
panel: analogous to the left but shows the 500-year events $\Delta_i
/ \sigma_t$ for SPA12, MOB and, as a further comparison, the
$\Delta$SSN$_i$/$\sigma_t$ (red) of the yearly sunspot number record
SSN (see paragraph 6 for relevance).}
\end{figure}
Figure 7 (left panel) depicts the F(s) and F$_2$(s) graphs of the FA
and DFA for SPA12 and MOB. The evaluated Hurst exponents are $\alpha
= \alpha _2 = 0.95$ $\pm$ 0.02 for SPA12 and $\alpha = \alpha _2 =
0.85$ $\pm$ 0.02 for MOB. The relevance for the further graphs in
the left panel of Figure 7 will be given in Eq. (12). In the case of
MOB, the Hurst exponent $\alpha_2$ from DFA has already been
evaluated as $ \alpha _2  = 0.86 \pm 0.02$ in \cite{Rybski2006}. The
right panel of Figure 7 shows the observed exceedance probabilities
W for $\Delta / \sigma_t$ $>$ 0 and (1-W) for $\Delta / \sigma_t$
$<$ 0 of both SPA12 and MOB together with the theoretical curve for
the Hurst exponent $\alpha$ = 0.95 derived from Eq. (11).
\begin{figure}[H]
\centering
\includegraphics[height=7.8cm]{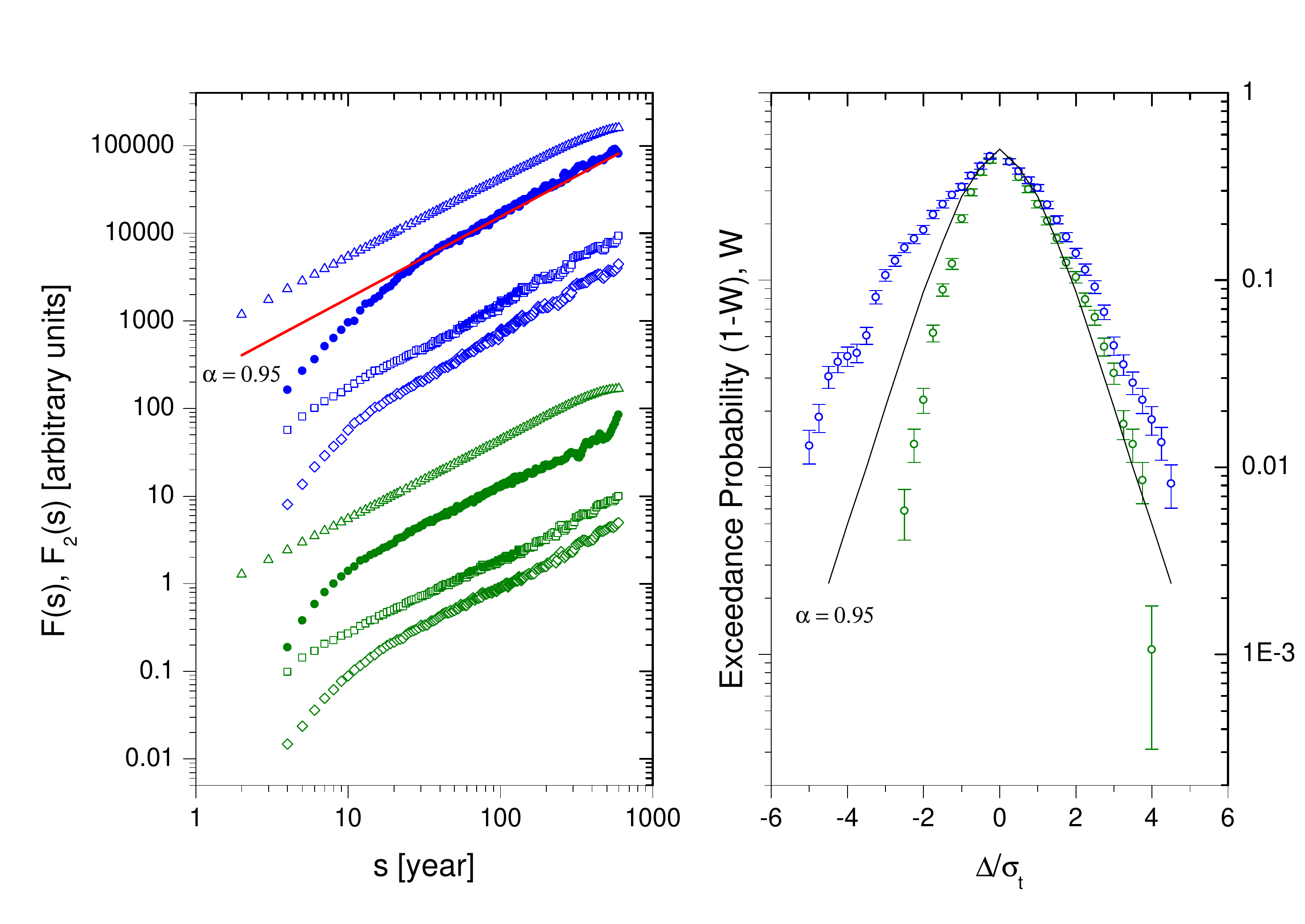}
\caption{(Color online) Left panel from top-down, with plotted
curves shifted for clarity; blue triangles: F(s) of the FA for
SPA12; blue filled circles: F2(s) of the DFA2 for SPA12; blue
squares: F2(s) of the DFA2 for a synthetic SPA12 record (see Eq.
(12) for relevance); blue diamonds: F2(s) of the DFA for a synthetic
SPA12 record with additional extreme short-term persistence. Below
this, the same graphs are repeated in green for MOB. The red line
with $\alpha$ = 0.95 is for comparison. Right panel: observed
exceedance probabilities for SPA12 (blue) and MOB (green) together
with the theoretical curve of Eq. (11) for $\alpha$ = 0.95. (1-W)
refer to negative values of $\Delta / \sigma_t$, W to positive ones.
For warming, the observed W values match roughly the theory, but the
cooling domain shows aberrations.}
\end{figure}
Because $ \alpha = \alpha _2$ has been evaluated for both SPA12 and
MOB, no linear trends have been removed by DFA. Obviously, the Hurst
exponents $ \alpha _2 \approx 0.9$ for the long-range records MOB
and SPA12 contradict their medium-range instrumental counterparts of
$ \alpha _2  \approx 0.6$ (see Table 2). As already stated, the
detrended Hurst exponents do not basically depend on the time scale
- months or years - and should therefore be similar for all
temperature records of sites located relatively close to each other.
In fact, all records in this paper - except for the MOB record - are
local, and the distance between them is no more than about 1000 km
at the most. Moreover, without further investigation into the
reliability of the long-range reconstructed records, the
discrepancies in the DFA results between the medium- and long-range
records cannot be explained. As Figure 7 demonstrates, the F$_2$(s)
graphs for SPA12 and MOB (filled circles) exhibit an unusual
downward dip for about s $ < $ 20 years that exceeds the weak
numerical effect already discussed. In an attempt to gain more
insight here, a simulation was carried out by synthetic records of
p$_i$ with a Hurst exponent $\alpha_2$ = 0.95, simulating SPA12, and
m$_i$ with $\alpha_2$ = 0.85, simulating MOB. As expected, the
F$_2$(s) graphs of both synthetic records p$_i$ and m$_i$ (open
squares in Figure 7) showed only the common weak downward dip. Next,
to simulate the strong downward dip, Eq. (12) was introduced to
furnish the additional conditions for p$_i$ and m$_i$:
\begin{equation}
\begin{array}{l}
 \bar p_{i + 2}  = p_i  + p_{i + 1}  + p_{i + 2} ,\;\;\;\;\;\;\;i = 3,......,1933 \\
 \bar m_{i + 2}  = m_i  + m_{i + 1}  + m_{i + 2} ,\;\;i = 3,......,1977 \\
 \end{array}
\end{equation}
This provides $\bar p_i$ and $\bar m_i$, which yield accordance
(open diamonds in Figure 7). An explanation of the strong downward
dip might be that the effective time resolution of both SPA12 and
MOB is in reality about 3-4 years, while the reconstructed
temperature for each year is a 3-4 year mean. A similar dip on
F$_2$(s) graphs in sea surface temperature (SST) records was found
and the hypothesis advanced that it has to do with the influence of
the North Atlantic Oscillation (NAO) and the El Ni$\tilde n$o
southern oscillation \cite{Monetti}.
\section{A hypothesis on the sun's influence}
Because no artifacts are known that could be responsible for a
strong persistence lasting at least 600 years in both tree ring and
stalagmite proxies (see left panel of Figure 7), another agent may
be supposed. For this purpose, a reconstructed sunspot number record
was used \cite{Solanki}. This covers more than 10,000 years (from
-9455 to 1895 AD) in 10-year time steps. Direct sunspot number
observations (Wolf numbers) as yearly averages for the period 1700 -
1995 AD provided additional support \cite{NOAA2005}. This completed
record has been given the name 'SSN' here, and covers a total of
11,450 years. It is used in smoothed 10-year steps. Annual and
monthly records are calculated from it by Spline interpolation. The
left panel of Figure 8 shows the SSN from 0 to 1995 AD. The right
panel depicts the DFA2 result for SSN, with 10 years steps and for
the whole record length of 11,450 years (smaller time steps for SSN
do not change the DFA2 result significantly). Apparently, the
F$_2$(s) curve for DFA2 in Eq. (8) does not follow the scaling law
of Eq. (5). As already stressed, there is an obvious conflict
between the Hurst exponents $\alpha _2 \approx 0.6$, derived from
the instrumental data (see Table 2 and Figure 4 in paragraph 4), and
$\alpha _2 \approx 0.9$, derived from the reconstructed temperatures
for SPA12 and MOB (see Figure 7 in paragraph 5). This paper advances
the hypothesis of fluctuations in sunspot numbers, which are
depicted in the left panel of Figure 8 as an agent here. The sun's
magnetic field modulates the cosmic ray flux hitting the earth. Both
cloud generation by cosmic rays and Earth temperatures are the
subject of intense discussion \cite{Kirkby2008}, \cite{Kirkby2011},
\cite{Krivova}, \cite{Lockwood2010}, \cite{Lockwood2011}, \cite{Lu},
\cite{Scafetta}, \cite{Shaviv}, \cite{Solanki}, \cite{Svensmark},
\cite{Weber2011}.

The reconstructed sunspot number record SSN in Figure 8 is a visual
re\-presentation of this presumed mechanism. The relatively slow
fluctuations in sunspot numbers - as opposed to the fast
fluctuations in monthly temperatures - act as a deterministic trend
over the last two hundred years that is roughly linear for the short
intervals of 20 to 50 years.
\begin{figure}[H]
\centering
\includegraphics[height=7.4cm]{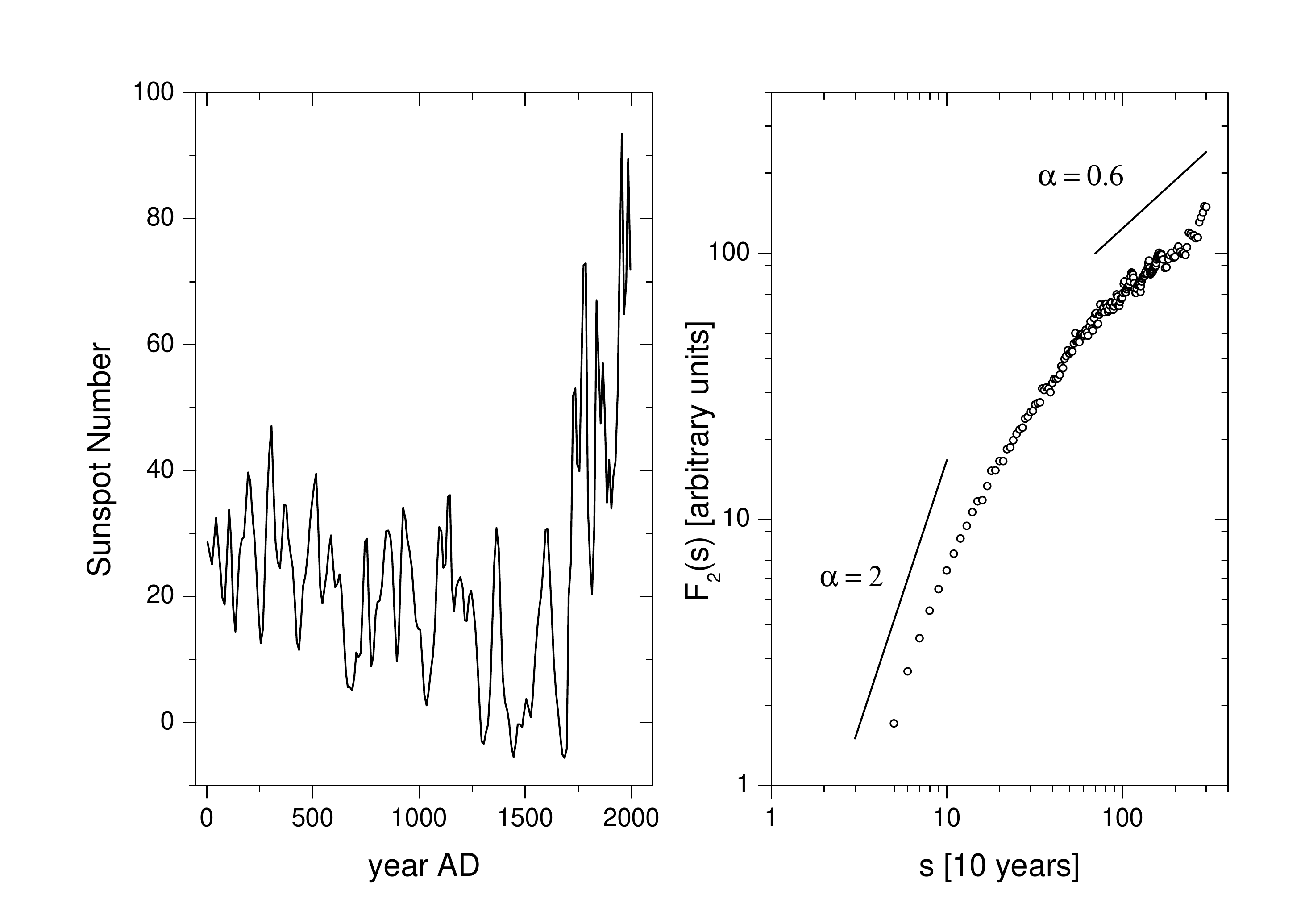}
\caption{(Left panel: sunspot number record SSN during the last 2000
years. Right panel: F$_2$(s) from DFA2 for the SSN sunspot number
record from -9455 AD to 1895 AD in 10-year steps. F$_2$(s) does not
take the usual scaling behaviour of $F(s) \sim s^{\alpha}$ of Eq.
(5) for long-term correlated records, but instead shows a sliding
gradient without any distinctive cross-over before s = 3000 years.
The s values up to about 100 years have an $\alpha _2  \approx 2$,
which drops down to $\alpha _2 \approx 0.6$ for s $ > $ 1000 years.
The lines with $\alpha$ = 2 and $\alpha$ = 0.6 are for comparison.}
\end{figure}
In the medium-range records, the SSN
trend is therefore identified as linear by DFA2 and is removed,
which explains the low detrended Hurst exponents. However, in the
long-range SPA12 and MOB records, the fluctuations in SSN are not
linear, and are, therefore, not removed by DFA2. Consequently, they
increase the Hurst exponent. The sunspot number hypothesis claims
that the instrumental records would also reveal greater Hurst
exponents provided that they were well over 250 years in length.
Unfortunately, no such records exist. This means that the hypothesis
could only be tested on synthetic records. A 17,880 month (1490
years) synthetic record x$_i$ with a Hurst exponent of $\alpha$ =
0.6, simulating the medium-range instrumental records, was generated
and superimposed on the monthly sunspot number record
\textit{ssn}$_i$. This \textit{ssn}$_i$ was established from the
part of SSN that corresponds to the time interval 505 - 1995 AD by
applying Spline interpolation and normalization. The superimposition
is carried out by means of
\begin{equation}
css_i (\xi ) = \left( {1 - \xi } \right)x_i  + \xi  \cdot ssn_i
\;\;\;\;i = 1,.....,17880\;\;[months]\;\;
\end{equation}
This results in the combined record \textit{css$_i$}($\zeta$) with a
temporarily unknown factor $\zeta$ and is referred to here as CSS.
\begin{figure}[H] \centering
\includegraphics[height=7.8cm]{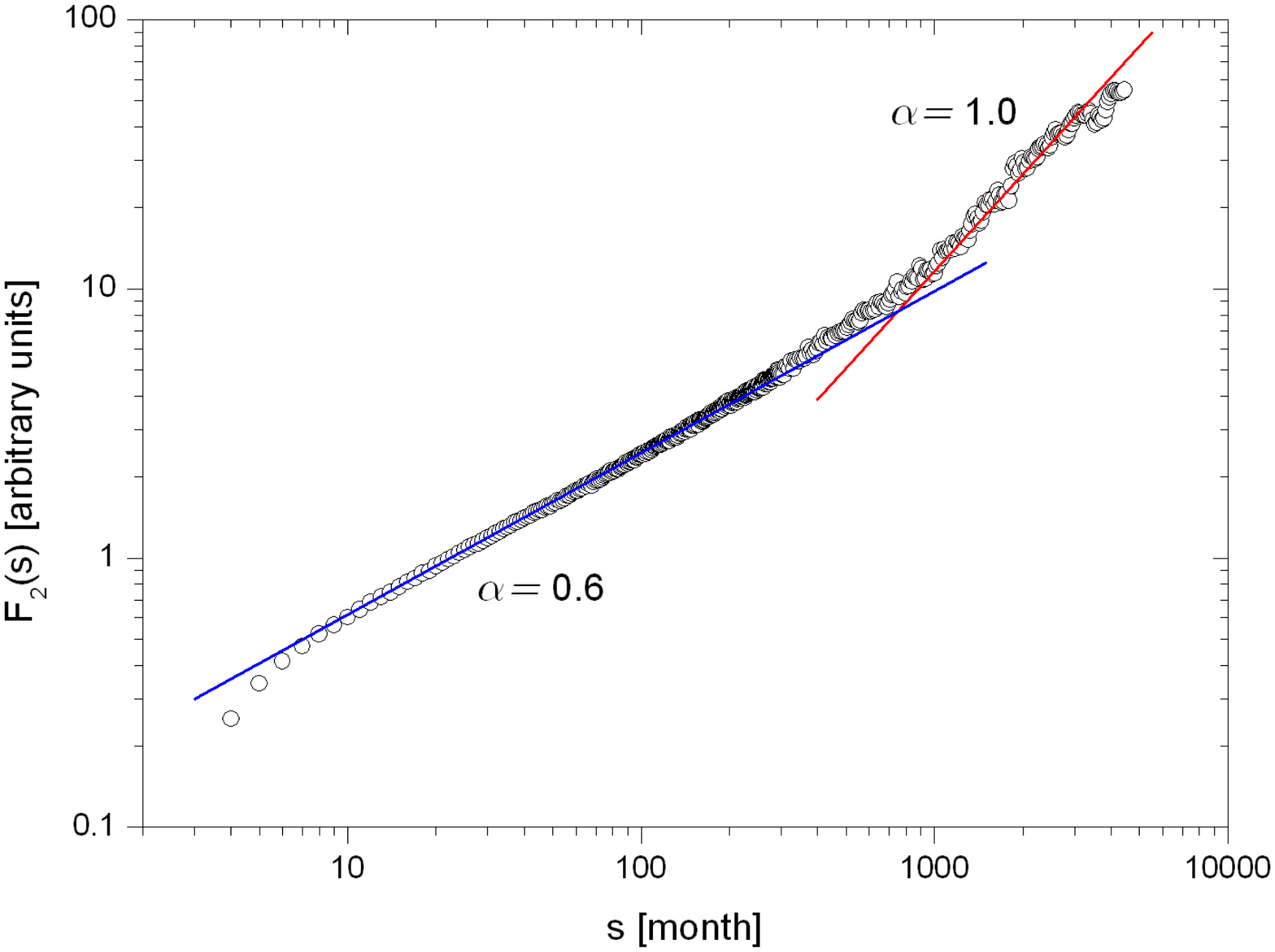}
\caption{(Color online) F$_2$(s) graph of the detrended fluctuation
analysis (DFA2) for the combined monthly record CSS of Eq. (13) over
a total period of 17,880 months (1490 years). The cross-over at
about 700 months ($\approx$ 60 years) indicates a lower limit for a
possible detection of the sun's influence by DFA2 in the
medium-range records. On the left of it, the Hurst exponent
$\alpha_2$ corresponds to the medium-range instrumental records at
$\alpha _2 \approx 0.6$ (see Figure 4 and Table 2 in paragraph 4)
and, on the right, to the long-range records SPA12 and MOB at
$\alpha _2 \approx 0.9$ (see Figure 8 in paragraph 5). Because the
reliability of DFA for monthly records of 200 years in length is
restricted to about 50 years, the instrumental records are too short
to reveal the cross-over.}
\end{figure}
The first month in Eq. (13) corresponds to the year 505 AD and i =
17,880 corresponds to 1995 AD. Eq. (13) follows the method of
superimposing trends on synthetic records \cite{Rybski2009}. Because
the sunspot hypothesis demands that CSS shows both the features of
the medium-range and long-range records, a value $\zeta$ = 0.013 was
identified that achieves this purpose, as Figure 9 demonstrates.
Figure 9 unveils a cross-over between 500 and 1000 months (40 - 80
years). On the left of it, the Hurst exponent $\alpha _2 \approx
0.6$ corresponds to the instrumental records and on the right,
$\alpha _2 \approx 1$ corresponds to SPA12 and MOB. It should be
pointed out that merely superimposing two synthetic records with
different Hurst exponents will not show a crossover. Figure
\nolinebreak9 provides an explanation of the different Hurst
exponents in the instrumental and reconstructed records: the
detrended fluctuation analysis DFA2 removes the sunspot fluctuations
as linear trends from the instrumental records that results in their
low Hurst exponents. It was already shown due to Eq. (9) that
F$_2$(s) is only feasible for s $\le$ N/4, hence for monthly records
that are about 220 years in length the limit of about s = 50 years
holds. As a consequence, series that are not distinctly over 250
years long are basically too short to reveal by DFA2 the extreme
long-term persistence caused by the sun. In Figure 4, in fact, no
cross over is visible, except for a rather feeble one for Paris.
However, in the case of the longer time frame, the situation
changes. If we look at the CSS synthetic record as well as at SPA12
and MOB, the SSN sunspot numbers correspond here to a fast
fluctuation, which increases the Hurst exponent $\alpha_2$ and is
not removed by DFA2.
\section{Conclusion}
Instrumental records going back a maximum of up to about 250 years
from present show the temperature declines in the 19$^{th}$ century
and the rises in the 20$^{th}$ to be of similar magnitudes. If we
assume anthropogenic CO$_2$ to be the agent behind the 20$^{th}$
century rise, we face a problem when it comes to the 19$^{th}$
century. The detrended fluctuation analysis (DFA2) evaluated - for
the five records selected here - very small natural probabilities
for both centennial events. Therefore, the probability that they are
the result of deterministic trends is high, but what actually caused
the trends to be diametrically opposed remains unknown. In contrast,
two high-quality long-range records, SPA12 and MOB, show frequent
centennial rises and falls of equal and greater magnitudes than
their shorter instrumental counterparts during the last 2000 years.
Smoothed SPA12 and MOB records are reported to be in accordance with
other biological proxies, indicating that centennial fluctuations at
least as strong as those of the past 250 years were indeed common
events. This is further confirmed by the DFA Hurst exponents of
$\alpha _2 \approx 0.9$ for SPA12 and MOB that are far higher than
the $\alpha _2 \approx 0.6$ of the instrumental records. As a
consequence, the impact of anthropogenic greenhouse gases is most
probably a minor effect and - in view of the 19$^{th}$  century
temperature fall of similar magnitude - not appropriate as an
authoritative explanation for any temperature rise in the northern
hemisphere during the 20$^{th}$  century.

Because no reliable explanation can be given for the conflict
between the different Hurst exponents and probabilities in the
instrumental and reconstructed records, a hypothesis of solar
influence (manifesting itself in long-term sunspot fluctuations)
could be put forward to explain the contradiction. A monthly
synthetic record covering about 1500 years and using a Hurst
exponent of $\alpha$ = 0.6, which corresponds to the instrumental
records was therefore superimposed on the trend of the sunspot
numbers. The DFA2 result for this combined record shows that it
embodies both the short persistence of the instrumental data and the
long persistence of the reconstructed data. The hypothesis expressed
here suggests that the sun could be predominantly responsible for
the 100-year-long rises and falls in temperature over the last 2000
years.\newpage
\section{Acknowledgements}
The author would like to thank S. Lennartz and A. Bunde (Department
of Theoretical Physics III, University of Giessen - Germany), A.
Mangini (University of Heidelberg and Academie of Sciences of
Heidelberg - Germany) and K.O. Greulich (Fritz Lipmann Institute,
University of Jena - Germany) for many helpful discussions, and
furthermore, L. Motl for his hint about the Czech source of the
Prague record.


\begin{thebibliography}{60}
%
\bibitem{Barnett}
T. Barnett et al., Detecting and attributing external influences on
the climate system: A review of recent advances, \textit{J. Clim.}
\textbf{18}(9), p. 1291 (2005)
%
\bibitem{Bogachev2007}
M.I. Bogachev, J.F. Eichner, and A. Bunde, Effect on Nonlinear
Correlations on the Statistics of Return Intervals in Multifractal
Data Sets, \textit{Phys. Rev. Lett.} \textbf{99}, 240601 (2007)
%
\bibitem{Bogachev2009}
M.I. Bogachev, I.S. Kireenkov, E.M. Nifontof, and A. Bunde,
Statistics of return intervals between long heartbeat intervals and
their usability for online prediction of disorders, \textit{New
Journal of Physics} \textbf{11}, 063036 (2009)
%
\bibitem{Bunde2001}
A. Bunde, J.W. Kantelhardt, Langzeitkorrelationen in der Natur: von
Klima, Erbgut und Herzrhythmus, \textit{Phys. Bl\"{a}tter}
\textbf{57}, Nr. 5 (2001)
%
\bibitem{Bunde2002}
A. Bunde, J. Kropp, and H.-J. Schellnhuber, The science of disaster:
Climate disruptions, market crashes and heart attacks,
\textit{Springer Verlag}, Berlin (2002)
%
\bibitem{Charney}
J.G. Charney, and J. G. de Vore, Multiple flow equilibra in the
atmosphere and blocking, \textit{J. Atmos. Sci.} \textbf{36}, p.
1205 (1979)
%
\bibitem{Eichner2003}
J.F. Eichner, E. Kosscielny-Bunde, A. Bunde, S. Havlin, and H.-J.
Schellnhuber, Power-law persistence and trends in the atmosphere: A
detailed study of long temperature records, \textit{Phys. Rev. E}
\textbf{68}(4), doi:10.1103/PhysRevE.68.046133 (2003)
%
\bibitem{Eichner2006}
J.F. Eichner, J.W. Kantelhardt, A. Bunde, and S. Havlin, Extreme
value statistics in records with long-term persistence,
\textit{Phys. Rev. E.} \textbf{73}, doi: 10.1103/PhysRevE.73.016130
(2006)
%
\bibitem{Esper}
J. Esper, E.R. Cook, and F.H. Schweingruber, Low-Frequency Signals
in Long Tree-Ring Chronologies for Reconstructing Past Temperature
Variability, \textit{Science} \textbf{295}, 2250, doi:
10.1126/science.1066208 (2002)
%
\bibitem{Fraedrich}
K. Fraedrich and R. Blender, Scaling of Atmosphere and Ocean
Temperature Correlations on Observations and Climate Models,
\textit{Phys. Rev. Lett.} \textbf{90}, 108501 (2003)
%
\bibitem{GISS}
http://data.giss.nasa.gov/gistemp/station\_data/
%
\bibitem{Hasselmann}
K. Hasselmann, Optimal Fingerprints for the detection of
time-dependent climate change, \textit{J. Clim.} \textbf{6}(10), p.
1957 (1993)
%
\bibitem{Hegerl}
G.C. Hegerl, H. von Storch, K. Hasselmann, B.D. Santer, U. Cubasch,
and P.D. Jones, Detecting greenhouse-gas-induced climate change with
an optimal fingerprint method, \textit{J. Clim.} \textbf{9}(10), p.
2281 (1996)
%
\bibitem{Hurst}
H.E. Hurst, Long-term storage capacity of reservoirs,
\textit{Transactions of the American Society of Civil Engineers}
\textbf{116} (2447) p. 770 (1951)
%
\bibitem{Ivanov}
P.C. Ivanov et al., Sleep-wake differences in scaling behaviour of
the human heartbeat: Analysis of terrestrial and long-term space
flight data, \textit{Europhysics Letters} \textbf{48}(5), p. 594
(1999)
%
\bibitem{Kantelhardt2001}
J.W. Kantelhardt, E. Koscielny-Bunde, H.H.A. Rego, S. Havlin, and A.
Bunde, Detecting long-range correlations with detrended fluctuation
analysis, \textit{Physica A} \textbf{295}, p. 441 (2001)
%
\bibitem{Kantelhardt2004}
J.W. Kantelhardt, Fluktuationen in komplexen Systemen (Fluctuations
in complex Systems), professorial dissertation, \textit{University
Giessen} (Germany), 19. June 2004
%
\bibitem{Kantelhardt2006}
J.W. Kantelhardt, E. Koscielny-Bunde, D. Rybski, P. Braun, A. Bunde,
S. Havlin, Long term persistence and multifractality of
precipitation and river runoff records, \textit{Journal of
Geophysical Research-Atmosphere} \textbf{111}, p. 1106 (2006)
%
\bibitem{Kiraly}
A. Kir�ly, I. Bartos, and I.M. J�Anosi, Correlation properties of
daily temperature anomalies over land, \textit{Tellus, Ser. A}
\textbf{58}(5), p. 593 (2006)
%
\bibitem{Kirkby2008}
J. Kirkby, Cosmic Rays and Climate, \textit{European Organisation
for Nuclear Research}, CERN-PH-EP/2008-005 (2008)
%
\bibitem{Kirkby2011}
J. Kirkby et al., Role of sulfure acid, ammonia and galactic cosmic
rays in atmospheric aerosol nucleation, \textit{nature}
\textbf{476}, p. 429, doi:10.1038/nature10343 (2011)
%
\bibitem{klementinum}
http://xmarinx.sweb.cz//KLEMENTINUM.xls
%
\bibitem{Koscielny}
E. Koscielny-Bunde, A. Bunde, S. Havlin, H.E. Roman, Y. Goldreich,
and H.-J. Schellnhuber, Indication of a universal persistence law
governing atmospheric variability, \textit{Phys. Rev. Lett.}
\textbf{81}(3), p. 729 (1998)
%
\bibitem{Krivova}
N.A. Krivova, and S.K. Solanki, Solar variability as an input to the
Earth's environment, \textit{International Solar Cycle Studies
(ISCS) Symposium, 23 - 28 June 2003, Tatransk� Lomnica, Slovak
Republic. Ed.: A. Wilson. ESA SP-535, Noordwijk: ESA Publications
Division}, ISBN 92-9092-845-X, 2003, p. 275
%
\bibitem{Kropp}
J. Kropp, and H.-J. Schellnhuber, In Extremis: Trends, Correlations,
and Extremes in Hydrology and Climate, \textit{Springer-Verlag},
Berlin (2010)
%
\bibitem{Lennartz2009a}
S. Lennartz, and A. Bunde, Trend Evaluation in Records with
Long-term Memory: Application to Global Warming, \textit{Geophys.
Res. Lett.} \textbf{36}, L16706, doi:10.1029/ 2009GL039516 (2009a)
%
\bibitem{Lennartz2009b}
S. Lennartz, S, and A. Bunde, Eliminating finite-size effects and
detecting the amount of white noise in short records with long-term
memory, \textit{Phys. Rev. Lett. E} \textbf{79}, p. 066101 (2009b)
%
\bibitem{Lennartz2011}
S. Lennartz, and A. Bunde, Distribution of natural trends in
long-term correlated records: A scaling approach, \textit{Phys. Rev.
E} \textbf{84}, 021129 (2011)
%
\bibitem{Livina2003}
V.N. Livina, Y. Ashkenazy, Z. Kizner, V. Strygin, A. Bunde, and S.
Havlin, A stochastic model of river discharge fluctuations,
\textit{Physica A} \textbf{330}, p. 283 (2003)
%
\bibitem{Livina2005}
V.N. Livina, S. Havlin, and A. Bunde, Memory in the Occurrence of
Earthquakes, \textit{Phys. Rev. Lett.} \textbf{95}, 208501 (2005)
%
\bibitem{Lockwood2010}
M. Lockwood, R.G. Harrison, T. Woollings, and S.K. Solanki, Are cold
winters in Europe associated with low solar activity?,
\textit{Environ. Res. Lett.} \textbf{5}, 024001, doi:10.1088/
1748-9326/5/2/024001 (2010)
%
\bibitem{Lockwood2011}
M. Lockwood, Solar influence on the Earth's climate, ISSI Workshop
''Observing \& Modelling Earth's Energy Flow", 11. Jan. 2011
%
\bibitem{Loehle}
C. Loehle, Climate change: detection and attribution of trends from
long-term geological data, \textit{Ecolog. Model.} \textbf{171},
433-450, doi:10.1016/j.ecolmodel.2003.08.013 (2004)
%
\bibitem{Luedecke2011}
H. L\"{u}decke, R. Link, and F.-K. Ewert, How natural is the recent
Centennial Warming? An Analysis of 2249 Surface Temperature Records,
\textit{International Journal of Modern Physics C}, to be published
in Oct. 2011.
%
\bibitem{Lu}
Q.-B. Lu, Cosmic-ray-driven-electron-induced reactions of
halogenated molecules adsorbed on ice surfaces: Implications for
atmospheric ozone depletion, \textit{Physics Reports},
doi:10.1016/j.physrep.2009.12.002 (2009)
%
\bibitem{Luterbacher}
J. Luterbacher, D. Dietrich, E. Xoplaki, M. Grosjean, and H. Wanner,
European Seasonal and Annual Temperature Variability, Trends, and
Extremes Since 1500, \textit{Science} \textbf{303}, p. 1499 (2004)
%
\bibitem{Lux}
F. Lux, M. Ausloos, Market fluctuations I: Scaling, multiscaling,
and their possible origins, in ''the science of disasters",
\textit{Springer-Verlag}, Berlin, p. 373, chapter 13 (2002)
%
\bibitem{Mangini2005}
A. Mangini, C. Sp\"{o}tl, and P. Verdes, Reconstruction of
temperature in the Central Alps during the past 2000 years from
$^{18}$O stalagmite record, \textit{Earth and Planetary Science
Letters} \textbf{235}, p. 741 (2005)
%
\bibitem{Mangini2007}
A. Mangini, P. Verdes, C. Sp\"{o}tl, D. Scholz, N. Vollweiler, and
B. Kromer, Persistent influence of the North Atlantic hydrography on
central European winter temperature during the last 9000 years,
\textit{Geophys. Res. Lett.} \textbf{34}, L02704, doi:
10.1029/2006GL028600, 2007
%
\bibitem{Mantegna}
R.N. Mantegna, H.E. Stanley, An Introduction to Econophysics:
Correlations and Complexity in Finance, \textit{Cambridge University
Press}, Cambridge (2000)
%
\bibitem{Moberg}
A. Moberg, D.M. Sonechkin, K. Holmgren, N.M. Datsenko, and W.
Karlen, Highly variable Northern Hemisphere temperatures
reconstructed from low-and high-resolution proxydata,
\textit{Letters to Nature} \textbf{433}, (7026), p. 613 (2005)
%
\bibitem{Monetti}
R.A. Monetti, S. Havlin, and A. Bunde, Long term persistence in the
sea surface temperature Fluctuations, \textit{Physica A}
\textbf{320}, p. 581 (2003)
%
\bibitem{NOAA2010}
NOAA (2010), http://www.ncdc.noaa.gov/paleo/pubs/pcn/
%
\bibitem{NOAA2005}
NOAA (2005), Solanki, S.K., et al. (2005), 11,000 Year Sunspot
Number Reconstruction, IGBP PAGES/World Data Center for
Paleoclimatology , Data Contribution Series \#2005-015, NOAA/NGDC
Paleoclimatology Program, Boulder CO, USA
%
\bibitem{Palmer}
T.N. Palmer, Predicting uncertainty in forecasts of weather and
climate, \textit{Rep. Prog. Phys.} \textbf{63}, p. 71 (2000)
%
\bibitem{Pelletier}
J.D. Pelletier, and D. L. Turcotte, Self-affine time series: II.
Applications and models, \textit{Adv. Geophys.} \textbf{40}, p. 91
(1999)
%
\bibitem{Rybski2006}
D. Rybski, A. Bunde, S. Havlin and H. von Storch, Long-term
persistence in climate and the detection problem, \textit{Geophys.
Res. Lett.} \textbf{33}, L06718, doi:10.1029/2005GL025591 (2006)
%
\bibitem{Rybski2008}
D. Rybski, A. Bunde, and H. von Storch, Long-term memory in
1000-year simulated temperature records, \textit{J. Geophys. Res.}
\textbf{113}, D02106, doi:10.1029/2007JD008568 (2008)
%
\bibitem{Rybski2009}
D. Rybski, and A. Bunde, On the detection of trends in long-term
correlated records, \textit{Physica A} \textbf{388}, p. 1687 (2009)
%
\bibitem{Scafetta} N. Scafetta, P. Grigolini, T. Imholt, J. Roberts,
and B. West, Solar turbulence in Earth's global and regional
temperature anomalies, \textit{Phys. Rev. E} \textbf{69}, 026303
(2004)
%
\bibitem{Shaviv}
N.J. Shaviv, J. Veizer, Celestial driver of Phanerozoic climate?,
\textit{GSA Today} \textbf{3}(7) (2003)
%
\bibitem{Solanki}
S.K. Solanki, I.G. Usoskin, B. Kromer, M. Sch\"{u}ssler and J. Beer,
An unusually active Sun during recent decades compared to the
previous 11,000 years, \textit{Nature} \textbf{431}, 7012, p. 1084
(2004)
%
\bibitem{Svensmark}
H. Svensmark, T. Bondo, and J. Svensmark, Cosmic ray decreases
affect atmospheric aerosols and clouds, \textit{Geophys. Res. Lett.}
\textbf{36}, L15101, doi:10.1029/2009GL038429 (2009)
%
\bibitem{Talkner}
P. Talkner, and R.O. Weber, Power spectrum and detrended Fluctuation
analysis: Application to daily temperatures, \textit{Phys. Rev. E}
\textbf{62}, 150 (2000)
%
\bibitem{Weber2001}
R.O. Weber, and P. Talkner, Spectra and correlations of climate data
from days to decades, \textit{J. Geophys. Res.} \textbf{106} (D17),
20, 131-20, 144 (2001)
%
\bibitem{Weber2011}
W. Weber, Strong Signature of the Active Sun in 100 Years of
Terrestrial Insolation Data, \textit{Annalen der Physik}
\textbf{522}, no. 6, p. 372, DOI 10.1002/andp201000019 (2010)
%
\bibitem{wetterzentrale}
http://www.wetterzentrale.de
%
\bibitem{Yamasaki}
K. Yamasaki, L. Muchnik, S. Havlin, A. Bunde, and H.E. Stanley,
Scaling and memory in volatility return intervals in financial
markets, \textit{PNAS} \textbf{102}, no. 26, p. 9424 (2005)
%
\bibitem{Zorita}
E. Zorita, T.F. Stocker and H. von Storch, How unusual is the recent
series of warm years?, \textit{Geophys.Res.Lett.} \textbf{35},
L24706, doi:10.1029/2008GL036228 (2008)
%
\bibitem{Zwiers}
F.J. Zwiers, The detection of climate change, in Anthropogenic
Climate Change, p. 163, \textit{Springer}, New York (1999)
%
\end{thebibliography}
\end{document}